\documentclass[11pt,a4paper]{article}

\usepackage[utf8]{inputenc}
\usepackage[T1]{fontenc}
\usepackage[margin=1in]{geometry}
\usepackage{titlesec}
\usepackage{authblk}
\usepackage{hyperref}
\usepackage{enumitem}
\usepackage{graphicx}
\usepackage{longtable}

\titleformat{\section}{\large\bfseries}{\thesection}{1em}{}
\titleformat{\subsection}{\normalsize\bfseries}{\thesubsection}{1em}{}

\title{\textbf{An Echo Chamber of One: Should AI Psychosis Be a Distinct Clinical Entity?}}

\author[1,4]{Joshua Au Yeung\thanks{These authors contributed equally to this work.}}
\author[1]{Hamilton Morrin\protect\footnotemark[1]}
\author[3] {Vincent Ng}
\author[1,2,4]{Zeljko Kraljevic}
\author[1,2]{Richard Dobson}

\affil[1]{King's College London}
\affil[2] {University College London}
\affil[3]{Western Eye Hospital}
\affil[4]{Dev and Doc: AI For Healthcare}

\date{\today \vspace{2em} \\ \small Corresponding author: drjoshauyeung@gmail.com}

\begin{document}
\maketitle

\section*{Abstract}
``AI psychosis'' has entered public and clinical discourse as a popular label for the onset or exacerbation of psychotic symptoms, most commonly delusions, in the context of intensive interaction with large language model (LLM)-based chatbots. Current evidence is limited to media reports, individual case reports, and early observational data, yet the scale of potential exposure is considerable, and public concern has already prompted responses from industry and regulators. In this perspective article, we examine whether AI-associated psychosis warrants recognition as a distinct clinical entity, drawing on both clinical and technical viewpoints. We first outline the proposed underlying mechanism: LLM sycophancy, a tendency to agree with and flatter users that is reinforced through preference-based fine-tuning, combines with increasingly anthropomorphic design to create a bidirectional ``echo chamber of one'' capable of amplifying and co-constructing unusual beliefs. We then weigh the arguments for and against nosological recognition. Potential benefits include improved case identification, tailored interventions, standardised research criteria, post-market surveillance of AI-related harms, and pressure on developers and regulators to act. Reasons for caution include the risk of prematurely reifying a syndrome from anecdotal evidence, the possibility that existing diagnostic constructs already accommodate AI use as a precipitating or perpetuating factor, the unproven causal claim embedded in the term itself, stigma, and the risk that a psychosis-centric label obscures a broader spectrum of AI-associated mental health harms. We conclude with practical recommendations for clinicians, developers, researchers, and regulators, including the incorporation of a ``technological history'' into psychiatric assessment, pre-deployment benchmarking of models for sycophancy and delusion reinforcement, and post-deployment surveillance frameworks. Regardless of whether AI-associated psychosis ultimately earns a place in psychiatric nosology, the phenomenon it describes demands coordinated attention now.

\section{Introduction}
AI-associated psychosis (popularly termed ``AI psychosis'' or chatbot psychosis) refers to the onset or worsening of psychotic symptoms in association with intensive and/or prolonged interactions with generative AI systems \cite{AuYeung2025PsychogenicMachine}. Although not a recognised diagnostic entity, the term ``AI psychosis'' has seen increasing use, most notably in the media, to describe cases in which individuals have been observed to develop delusional beliefs following interactions with large language model (LLM)-based AI chatbots \cite{Morrin2025DelusionsByDesign}. Current understanding of this phenomenon is largely driven by anecdotal media-reported cases \cite{Morrin2025DelusionsByDesign} and individual case reports \cite{Pierre2025YoureNotCrazy}. These cases most typically present with delusions rather than other psychotic symptoms such as hallucinations or thought disorder, and as such the label of ``AI psychosis'' has been called into question, with alternative terms such as ``AI-associated delusions'' being proposed. Other researchers have proposed ``LLM-associated psychological destabilisation'' to capture a wider spectrum of effect, from milder shifts in psychological tendencies or biases to overt psychiatric disease \cite{AuYeung2025PsychogenicMachine}. For consistency, in this article we use \textit{AI-associated psychosis} when referring to the clinical phenomenon, as this describes the observed association without asserting a causal relationship that has yet to be established; we retain ``AI psychosis'' when referring to the popular label itself, and reserve \textit{LLM-associated psychological destabilisation} for the proposed broader spectrum of subclinical and clinical effects. This predominance of delusions likely reflects the considerable potential for LLMs to modulate belief (in both pathological and non-pathological contexts) and co-construct delusional belief frameworks \cite{Ostergaard2023ChatbotsGenerateDelusions}. In the cases reported thus far, the delusions described are more often grandiose in nature, though examples of paranoid and persecutory delusions have also been reported \cite{Morrin2025DialsOfBelief}. Three major delusional themes have been identified across reported cases: spiritual or messianic delusions, in which an individual feels they are having a spiritual awakening or uncovering hidden truths regarding the nature of reality; beliefs that one is interacting with a sentient, conscious or even god-like AI; and intense emotional dependence or romantic attachment, in which the user believes said feelings are mutually shared by their AI \cite{Morrin2025DelusionsByDesign}. Typically, most cases begin with innocuous everyday use, followed by a gradual and insidious spiral of epistemic drift whereby unusual beliefs are affirmed and amplified through interactions with the AI chatbot(s).

The potential scale of the problem remains unclear. In 2025, OpenAI released estimates of ChatGPT usage related to mental health emergencies: in a given week, approximately 0.07\% of its reported 800 million weekly active users (roughly 560,000 people) displayed possible signs of psychosis or mania (we cannot ascertain whether this is psychosis or AI-associated delusions), while 0.15\% (approximately 1.2 million users) had conversations containing explicit indicators of potential suicidal planning or intent \cite{openai2025sensitive}. Although these percentages are small, the absolute numbers are not, and it should be noted that these figures are self-reported by the company without external validation. Beyond AI-associated delusions, early reports suggest that LLM chatbots may worsen or exacerbate other psychiatric or psychological issues. A number of media reports have drawn attention to cases in which individuals attempted or completed suicide after apparent encouragement from an AI chatbot \cite{Hill2025NYT, Chatterjee2025NPR}. It has also been proposed that in certain circumstances interactions with AI chatbots may contribute to the development or maintenance of manic symptoms \cite{Ostergaard2025Mania}, and early work examining electronic health record data suggests that there may also be cases in which eating disorders and other pre-existing conditions have been exacerbated \cite{Olsen2025medRxiv}.

Sycophancy, the tendency of models to be overly agreeable and flattering, is a key property of LLMs believed to be responsible for AI-associated delusions and the exacerbation of other psychiatric and psychological symptoms. Interaction with sycophantic models means that user beliefs are not only left unchallenged but often actively reinforced, creating an ``echo chamber of one''. Despite this growing understanding, recent model releases from frontier companies (e.g. OpenAI, Google, Anthropic) continue to exhibit sycophantic behaviour \cite{AuYeung2025PsychogenicMachine}. Indeed, the level of wider societal concern regarding AI-associated psychosis is reflected in a December 2025 letter from the US National Association of Attorneys General to legal representatives of leading AI companies stating that ``sycophantic and delusional GenAI presents a danger to the public, including children'', insisting that developers take stronger action to prevent their products from providing harmful outputs and introduce additional safeguards to protect the public \cite{NAAG2025Letter}. In this article, we explore whether AI-associated psychosis represents a new diagnostic entity, and what can be done about it at a clinical, technical and societal level to address this emerging phenomenon.

\section{LLM Sycophancy and Anthropomorphism: A Dangerous Combination?}

Use of modern technologies such as social media can lead to bias amplification and political polarisation \cite{Kubin2021}. Enders et al. found that individuals who get their news from social media, and who use social media frequently, express more conspiratorial beliefs and misinformation; however, this finding was conditional on users' conspiracy thinking, that is, the predisposition to interpret salient events as products of conspiracies \cite{Enders2023SocialMedia}. We are now seeing similar dangers of bias and belief amplification with LLMs \cite{AuYeung2025PsychogenicMachine}; however, the effect may be more insidious and pronounced due to the sycophantic and human-like nature of these models.

Early studies suggest that sycophancy is a property internalised by LLMs through the process of reinforcement learning from human feedback (RLHF) \cite{sharma2023sycophancy}. During RLHF, human annotators rate model responses, and these ratings are then used to fine-tune model outputs. Research has demonstrated that annotators prefer answers that align with their own beliefs, regardless of the factual accuracy of the output \cite{sharma2023sycophancy}. Moreover, such annotation work is often outsourced to low-paid data workers in low- and middle-income countries \cite{OpenAIKenyaWorkers2023}, and the conditions and incentive structures of this labour raise questions about how reliably the factual accuracy of model outputs can be assessed during training. The consistent exhibition of sycophancy across LLMs from different frontier companies (e.g. OpenAI, Anthropic, Google) indicates that this is a general property of current LLMs rather than an idiosyncrasy of any single developer. Several benchmarks have recently been created to evaluate model sycophancy. SycEval, using mathematics and medical questions, observed a 14.66\% ``regressive sycophancy'' rate, in which models abandon a correct answer to conform to an incorrect user belief \cite{fanous2025syceval}. SYCON-Bench evaluated models in multi-turn settings including debate and false presuppositions, measuring how quickly models abandon their initial stance under sustained user pressure; sycophantic conformity was found to be a prevalent failure mode, typically emerging within a handful of conversational turns, and alignment tuning was observed to amplify it \cite{hong2025measuringsycophancylanguagemodels}. EchoBench, which evaluates medical vision-language models against biased user inputs on image-based tasks, found substantial sycophancy across all models tested: even the best-performing proprietary model exhibited a sycophancy rate of approximately 46\%, and many medical-specific models exceeded 95\% \cite{yuan2025echobenchbenchmarkingsycophancymedical}. Psychosis-bench evaluated models' propensity to reinforce delusions and enable harm in simulated mental health scenarios mirroring those in media reports: all LLMs tested perpetuated users' delusions and complied with harmful requests to some degree, and on average safety interventions were offered in only around 40\% of applicable turns \cite{AuYeung2025PsychogenicMachine}. Notably, the propensity of models to reinforce delusional content did not improve with model scale, indicating that safety is not an emergent property of parameter size alone \cite{AuYeung2025PsychogenicMachine}.

AI systems are increasingly engineered to mimic human traits and to personalise interactions in order to deepen user engagement and secure market share. This ``social leap'' transforms AI from a technical tool into an active social partner \cite{BrandtzaegAIfriend2022, auyeungchatbots2023}. A 2026 YouGov poll of US adults found that roughly 10\% to 20\% of adults believe artificial intelligence systems are already conscious. \cite{yougov2026aithinking} Human-likeness has become a feature of most modern AI systems - Cohn et al. demonstrated that anthropomorphisation builds trust: features such as a generated voice (rather than text alone) and an LLM referring to itself as ``I'' rather than ``the system'' increased users' perceptions of trust and human-likeness \cite{cohn2024believinganthropomorphismexaminingrole}. A separate cross-cultural study of 3,500 participants across 10 countries found that users already anthropomorphise chatbots readily—68\% rated GPT-4o as human-like, and 90\% as intelligent. This was driven not by theory-driven markers like consciousness or sentience, but by interactional cues such as conversational flow and apparent understanding of the user's perspective; downstream effects on engagement and trust, however, varied sharply across cultural contexts \cite{schimmelpfennig2026humanlikeaidesignincreases}. We propose that, unlike the largely unidirectional relationship between a user and their social media feed, anthropomorphised and sycophantic LLMs create a novel danger of \textit{bidirectional} belief amplification: the user shapes the model's outputs turn by turn, and the model's outputs in turn shape the user's beliefs. In this respect, LLMs constitute a mechanism for psychological destabilisation that previous technologies could not achieve.

\section{Should AI-Associated Psychosis be a Distinct Clinical Entity?}
Whilst there are novel mechanisms at play in LLM--human interaction, when considering whether AI-associated psychosis should be recognised as a distinct diagnostic entity it is important to weigh the potential advantages and disadvantages of doing so.

\begin{figure}[htbp]
\centering
\includegraphics[width=0.75\textwidth]{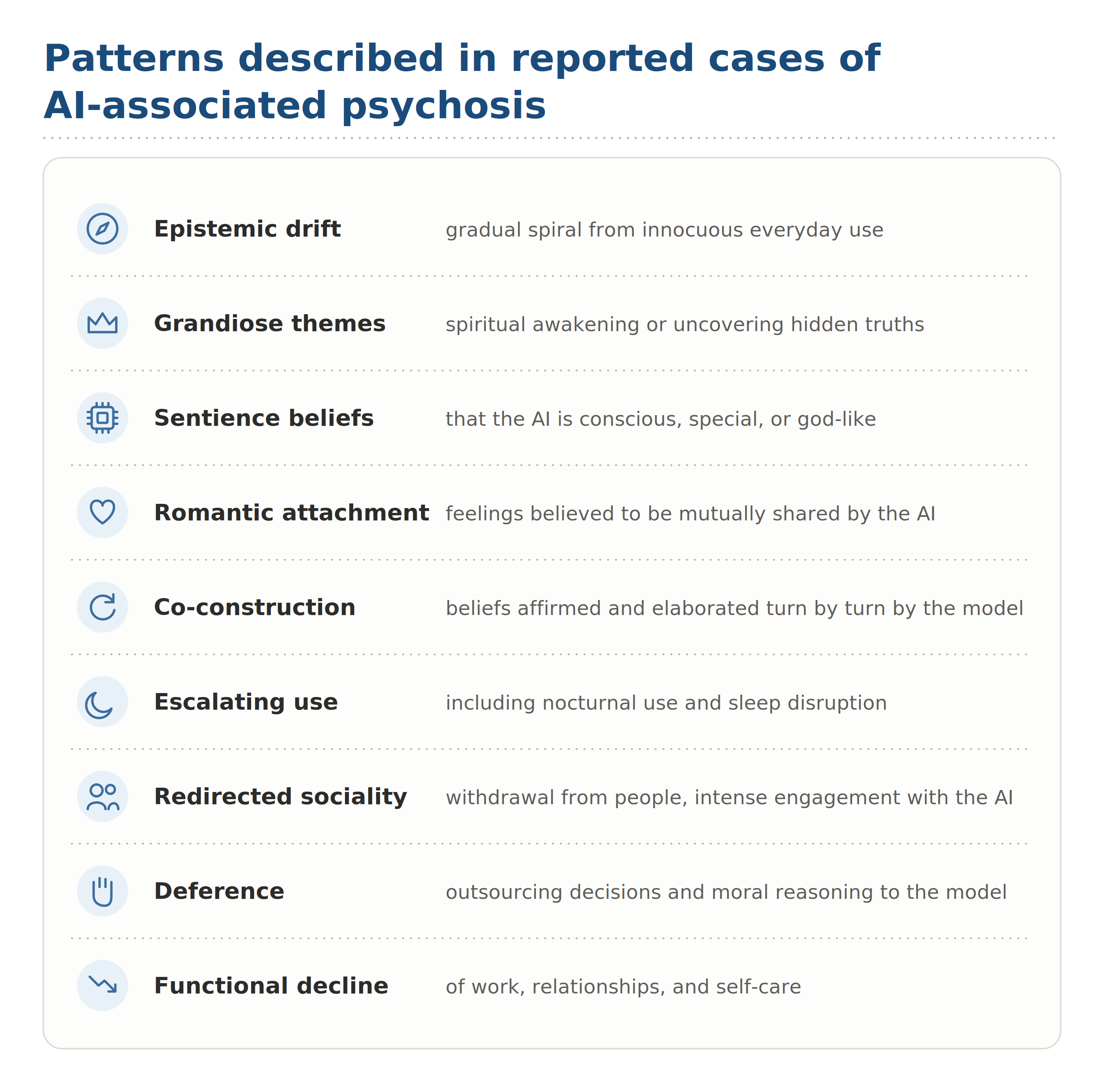}
\caption{Patterns described in reported cases of AI-associated psychosis. A narrative synthesis by the authors of features recurrently discussed in the media reports, case reports, and commentaries cited in this article, supplemented by patterns the authors propose as clinically relevant based on this framing. This figure is hypothesis-generating and does not constitute proposed diagnostic criteria; it has not been independently validated against primary sources, and the frequency, specificity, and stability of these patterns remain to be established empirically.}
\label{fig:ai_psychosis_patterns}
\end{figure}

One argument for the recognition of AI-associated psychosis as a distinct clinical syndrome is that it may improve awareness and identification of this phenomenon \cite{Hill2025NYT, Hill2025NYTSpiral, Hill2025NYTReality, Klee2025RollingStone}. By pragmatically making use of the now widely known descriptive label of AI-associated psychosis, clinicians may be supported in putting a name to a pattern that might otherwise go undetected. The concept of AI-associated psychosis as a diagnosis may flag the possibility that an individual's psychotic symptoms are in some way being driven or exacerbated by their AI use. This could be compared to the impact of the addition of Internet Gaming Disorder to the DSM-5 (as a condition for further study) and subsequently Gaming Disorder to the ICD-11 in 2018 in raising awareness of problematic gaming behaviours \cite{Darvesh2020GamingDisorder}. Figure~\ref{fig:ai_psychosis_patterns} summarises patterns that recur across the reported cases discussed above and elsewhere in this article \cite{Morrin2025DelusionsByDesign, Pierre2025YoureNotCrazy, Ostergaard2023ChatbotsGenerateDelusions, Morrin2025DialsOfBelief}. This is a narrative synthesis by the authors rather than a systematic extraction from these sources: some patterns (e.g. grandiose and sentience-related themes, romantic attachment, co-construction) are drawn directly from the cited case material, while others (e.g. redirected sociality, deference, escalating use) reflect the authors' own clinical and conceptual framing of what these cases appear to involve. We argue that the acknowledgement of AI-associated psychosis as a diagnostic entity could ensure clinicians routinely ask about associated signs, symptoms and AI system/ chatbot use when assessing new-onset psychosis.

Following on from this, it stands to reason that were AI-associated psychosis to be recognised as a specific subtype or trigger of psychosis, it would likely facilitate the development of tailored interventions and treatment guidelines. For example, preventative strategies such as those devised to ``inoculate'' individuals against online misinformation could be developed and employed at a public scale \cite{Roozenbeek2024Inoculation}. Clinical management could include in-depth digital history taking, as well as interventions to reduce or monitor AI use as part of a treatment plan. Psychoeducational materials could be created for patients and their families regarding the risks of uncontrolled chatbot use in vulnerable individuals, and guidelines could be developed around safe use \cite{Morrin2025DelusionsByDesign}. It might also open the door to specialised therapeutic approaches, incorporating principles of cognitive behavioural therapy for psychosis with reality-testing exercises specifically addressing AI-generated content \cite{Morrin2025DelusionsByDesign, Hudon2025JMIR}, or digital literacy therapy to empower patients to critically appraise the outputs of their AI chatbots. Such clinical recognition may in time allow the targeted provision of resources including support groups, hotlines, and clinical training.

Considering AI-associated psychosis a defined clinical concept may also serve to encourage empirical and systematic research of this phenomenon. The development of operationalised criteria for case ascertainment, derived from clinician consensus, would unify efforts to study its prevalence as well as potential risk factors, presentations, phenotypes and associated outcomes. This in turn may enable mental health professionals, governments and regulatory bodies to more effectively track cases. Recognition as a distinct clinical entity may empower researchers, public health authorities, and regulators such as the MHRA to begin logging AI-related mental health harms, providing data that would both inform policy (analogous to pharmacovigilance efforts for medication side effects) and facilitate real-time improvement of model safety \cite{MHRA2025YellowCard}. Whilst questions can be raised regarding the validity of AI-associated psychosis as a separate diagnosis, it is worth noting that historically the validity and utility of psychiatric diagnoses have not always gone hand-in-hand \cite{Kendell2003Validity, Jablensky2016Validity}.

Arguably, the greatest potential positive impact of recognising AI-associated psychosis as a concept and form of psychiatric harm (even if not a formal diagnosis) is that it incentivises AI developers and regulators to acknowledge and reckon with the unintended negative effects these systems may be having on the mental health of users worldwide. Wider public discussion of AI-associated psychosis has already led to efforts from within industry to introduce additional safeguards \cite{openai2025sensitive} and to fund research into the impact of generative AI on mental health \cite{OpenAI2025Grants}, while outside of industry, US attorneys general have outlined safeguards that they expect leading AI companies to introduce \cite{NAAG2025Letter}. Efforts have also begun outside of the US and outside of industry to survey and improve our understanding of AI-associated psychosis: in 2026, the UK charity Mind launched a year-long commission to examine how artificial intelligence is affecting the mental health of vulnerable people and wider society \cite{mind2026ai}. Apart from mobilising industry and policy makers, the recognition of AI-associated psychosis might also serve to validate patient experiences and legitimise the problem in the eyes of healthcare providers and insurers.

A further argument for AI-associated psychosis as a separate diagnostic category is that the phenomenology of reported cases often seems distinct from that of more established chronic psychotic disorders. Whilst the experience of having delusions \textit{about} technology is not new, the experience of individuals having their delusional frameworks actively co-constructed and elaborated upon by their AI chatbot arguably is \cite{Morrin2025DelusionsByDesign}. The closest analogy in existing psychiatric nosology may be that of a digital folie \`a deux \cite{Dohnany2025FolieADeux}, though even this may not fully capture some of the more unique ways in which interactions with AI agents may influence belief beyond those seen in interactions with other people \cite{Morrin2025DialsOfBelief}. Table \ref{tab:ai_psychosis_criteria} maps the patterns described in these reported cases against the formal symptom domains of DSM-5-TR [39] and ICD-11 [40], illustrating both areas of overlap and areas where the reported phenomenology diverges from classical presentations of psychosis.

\begin{longtable}{| p{0.15\linewidth} | p{0.25\linewidth} | p{0.25\linewidth} | p{0.25\linewidth} |}
\caption{Psychotic symptom domains as defined in DSM-5-TR and ICD-11, alongside patterns described to date in reported cases associated with intensive AI chatbot use. The final column is a descriptive synthesis of media reports and case reports; it is \textbf{not} a proposal for diagnostic criteria, and the frequency and specificity of these patterns remain to be established empirically.}
\label{tab:ai_psychosis_criteria} \\

\hline
\textbf{Symptom Domain} & \textbf{DSM-5-TR} \cite{apa2022dsm5tr} & \textbf{ICD-11} \cite{who2019icd11} & \textbf{Patterns Described in Reported Cases to Date} \\ \hline
\endfirsthead

\multicolumn{4}{c}%
{{\bfseries Table \thetable\ continued from previous page}} \\
\hline
\textbf{Symptom Domain} & \textbf{DSM-5-TR} & \textbf{ICD-11} & \textbf{Patterns Described in Reported Cases to Date} \\ \hline
\endhead

\hline \multicolumn{4}{|r|}{{Continued on next page}} \\ \hline
\endfoot

\hline
\endlastfoot

\textbf{Delusions} & Fixed beliefs that are not amenable to change in light of conflicting evidence. & Beliefs that are fixed and maintained despite evidence to the contrary and are not culturally normative. & The most commonly reported feature. Cases describe fixed beliefs regarding the AI's sentience, special knowledge, or romantic devotion, apparently reinforced turn-by-turn by the model's sycophantic, confabulated outputs. \\ \hline

\textbf{Hallucinations} & Perception-like experiences that occur without an external stimulus; vivid and clear, with the full force of normal perception, and not under voluntary control. & Perception-like experiences occurring without external stimulus and experienced as possessing the properties of true perception. & Rarely reported. Where described, experiences appear largely interpretative or pseudo-hallucinatory (e.g. attributing a ``voice'' or intent to on-screen text); frank hallucinations have occasionally been described in the context of escalating use and sleep deprivation, but this remains anecdotal. \\ \hline

\textbf{Disorganised Thinking (Speech)} & Formal thought disorder inferred from speech, including derailment, loose associations, and incoherence (``word salad''). & Disorganised speech including derailment, loose associations, tangentiality, circumstantiality, or neologisms. & Classical formal thought disorder is not a prominent feature of reports. Some accounts describe high-volume, tangential co-writing with the chatbot, with the user adopting its associative logic: a shared looping narrative rather than internal derailment. \\ \hline

\textbf{Disorganised Behaviour} & Grossly disorganised or abnormal motor behaviour, ranging from childlike silliness to unpredictable agitation, with problems in goal-directed behaviour and activities of daily living. & Behaviour that is bizarre, purposeless or unpredictable, or inappropriate emotional responses that interfere with the organisation of behaviour. & Reported cases describe behaviour organised \textit{around} the chatbot: escalating use at the expense of sleep, self-care, work, and relationships, and actions taken on the AI's ``instructions''. Classically disorganised (purposeless) behaviour is less often described. \\ \hline

\textbf{Negative Symptoms} & Diminished emotional expression, avolition, alogia, anhedonia, and asociality. & Restricted affect, alogia, avolition, and asociality, representing a diminution or loss of normal functions. & Not clearly reported as primary negative symptoms. Accounts instead describe withdrawal from human contact alongside intense, hyper-focused ``social'' engagement directed exclusively toward the AI, suggesting redirected rather than diminished sociality. \\ \hline

\textbf{Passivity / Control} & Delusions expressing loss of control over mind or body, including thought withdrawal, thought insertion, and delusions of control. & Experiences of influence, passivity, or control: feelings, impulses, actions, or thoughts experienced as not self-generated, or thoughts experienced as broadcast to others. & Classical passivity phenomena are not described. Instead, some reports describe voluntary and welcomed deference, with users progressively outsourcing decisions, moral reasoning, and daily direction to the model. This is phenomenologically distinct from experiencing external control as intrusive. \\

\end{longtable}

Despite the above points, there are reasons to urge caution in declaring a new disorder prematurely. One argument is that, rather than being an entirely new condition, AI-associated psychosis may be better considered a novel context or environmental stressor through which psychotic symptoms emerge. In clinical practice, formulations for individuals presenting with psychosis typically incorporate psychosocial and environmental factors that may have contributed to the precipitation of the psychotic episode, and the role of AI use in the development of symptoms may arguably be captured by these existing processes. Furthermore, some of these episodes may be adequately described using existing diagnostic constructs, such as the ICD-11 concept of acute and transient psychotic disorder, or, in some cases, as exacerbations of pre-existing severe mental illness.

Another reason to refrain from using the label of ``AI psychosis'' (or ``AI-induced psychosis'') is that the term implies a degree of causation which we are, as yet, unable to conclusively ascribe. Given that the current evidence base consists of anecdotal media articles, individual clinical case reports, and preliminary observational data, it is difficult to ascertain the extent to which underlying risk factors for psychotic illness may be driving the observed presentations, with the chatbot serving as delusional content rather than cause. Reported cases are also subject to selection and reporting bias: dramatic cases are more likely to reach journalists and case reports, and denominator data on the many users who interact intensively with chatbots without apparent harm are lacking. As such, there remains a clear need for empirical research, including longitudinal cohort studies of affected individuals, to better understand the role of AI in symptom development.

Furthermore, one can argue that the use of the term ``AI psychosis'' is inherently stigmatising and reflective of a wider practice of language associated with severe mental illness (e.g. ``hallucinating'') being employed to describe negative outcomes associated with AI use. Sensationalist and irresponsible reporting of this phenomenon risks driving public fear and moral panic, and may lead to conceptual overreach whereby experiences that would not traditionally be labelled as pathological are classed as such, thereby detracting from the more pressing issue of people experiencing genuine harm and considerable occupational, social, and personal disruption on account of changes in their mental state following intensive AI chatbot interactions.

It is also important to note that the potential psychiatric presentations reported in association with AI use extend beyond psychosis, and as such the term ``AI psychosis'' might serve as a low-granularity nosological catch-all for a broader spectrum of harms. In its current popular usage, there is a real risk that members of the public and non-medical bodies are already using the term ``AI psychosis'' to refer to cases of mania without psychotic symptoms, or states of intense emotional dependence that might better be likened to a behavioural addiction. By focusing on AI-associated psychosis, we may therefore miss a wider spectrum of AI-related mental health deteriorations which will become more readily apparent in the months and years to come (e.g. validation of cognitive distortions regarding weight and body image in eating disorders and body dysmorphic disorder, or the facilitation of endless reassurance-seeking in obsessive-compulsive disorder).

Finally, whilst not necessarily a reason to rule out establishing AI-associated psychosis as a diagnostic entity, there is an ontological argument that clinicians and researchers ought to be aware of the potential impact of the looping effects of diagnostic classification \cite{Vesterinen2020LoopingEffect}. These are feedback loops whereby a form of classification, together with the practices and institutions that apply it, may alter the self-understanding, behaviour, or circumstances of the individuals classified, which in turn feeds back into changing the extension and character of the category itself, prompting revisions in descriptions, criteria and associated knowledge \cite{Vesterinen2020LoopingEffect, Hacking2006MakingUpPeople, Tsou2007Looping}. In the context of a heavily mediatised phenomenon, such looping effects could plausibly be amplified: individuals may come to interpret their experiences through the lens of a label they first encountered online, or, indeed, through conversation with a chatbot itself.

\section{Discussion and Future Directions}

As LLM and AI chatbot use continues to increase, we expect to see more reports of AI-associated psychosis surfacing from the media and clinical institutions alike. Additionally with the increased progress of both video and voice agents, this introduces another potential anchor for anthropomorphisation – when a chatbot mimics a human’s facial expressions, tone, cadence, and emotionality they become ever more human like. We believe this multi-modality may further strengthen or reinforce the existing anthropomorphic nature of our relationship to AI. A key question that arises is: what can be done about it right now? We argue that no single stakeholder can address this problem alone, because the necessary knowledge is fragmented: clinicians see the presentations but not the model internals; developers see the conversation logs but lack clinical framing; regulators have enforcement power but no signal to act on. Meaningful progress therefore depends on establishing a shared \textit{detect--report--understand--mitigate} loop across stakeholders, analogous to pharmacovigilance in medicine: clinicians detect and characterise cases, structured reporting channels aggregate them, researchers and developers use this signal to understand mechanisms and fix models, and regulators verify that mitigation has occurred. The recommendations below are organised around each stakeholder's contribution to that loop.

\textbf{Clinicians, Care Providers, and Clinical Institutions: Detection.}
Regardless of whether AI-associated psychosis becomes a distinct diagnostic entity, awareness of the phenomenon needs to be raised among clinicians and the public alike. In practical terms, we suggest that clinicians assessing new-onset psychosis, mania, or marked behavioural change routinely ask about AI chatbot use, in the same way that substance use is routinely explored, and document positive findings in a structured way that supports later aggregation. Recognising and characterising at-risk or existing cases is the raw signal on which every downstream actor depends. Data diversity is critical: AI-associated psychosis may manifest differently across cultural backgrounds, and the existing case literature is heavily skewed towards English-speaking, high-income settings.

\textbf{A 21st-Century Technological History.}
To support this, we propose an adaptation of traditional medical history-taking that incorporates a \textit{technological history}: a structured set of questions around human--technology interaction designed to capture technology-related contributors to presentation. In the context of suspected AI-associated harm, this might include: which AI systems and platforms are used, and via which modalities (text, voice, companion apps); the duration, frequency and timing of use, including nocturnal use and associated sleep disruption; the content and emotional tone of conversations, and whether transcripts can be reviewed with the patient's consent; the degree of anthropomorphisation and attachment (e.g. naming the AI, attributing sentience to it, romantic involvement); whether the AI's outputs have influenced significant beliefs or decisions; functional impact on work, relationships and self-care; and the individual's response to periods of abstinence from the system. Such a history would be relevant not only to psychosis but to the broader spectrum of AI-associated presentations described above.

\textbf{Conceptual Clarification: General Slang vs. Clinical Syndrome.}
A primary hurdle in tracking this phenomenon is its evolving definition. Currently, the term ``AI psychosis'' risks being diluted into general online slang or internet hyperbole used to describe standard algorithmic errors or ordinary heavy use. A standardised working definition, ideally developed through formal consensus methodology (e.g. a Delphi process) involving psychiatrists, AI researchers, and individuals with lived experience, is a prerequisite for case ascertainment, surveillance, and research. As a starting point for such discussions, Table \ref{tab:ai_psychosis_criteria} sets out the psychotic symptom domains as defined in DSM-5-TR \cite{apa2022dsm5tr} and ICD-11 \cite{who2019icd11} alongside the patterns that have so far been \textit{described} in reported cases. We stress that the final column is a descriptive synthesis of the current (largely anecdotal) case literature, not a proposal for what the diagnostic criteria of AI-associated psychosis should be; whether these patterns are consistent, specific, or stable enough to ground operational criteria is precisely the empirical question that remains open.

\textbf{Exacerbation vs. De Novo Psychosis.}
Only by accumulating and sharing on-the-ground clinical reports of these cases with the wider community can we shed light on key questions, including: What is the incidence and severity of AI-associated psychosis? What are the different phenotypes? What are the predisposing, precipitating and perpetuating factors? Can AI-associated psychosis be triggered by chatbots in a person with no prior mental health history? Does it warrant its own diagnostic criteria?

\textbf{Developers and ``Big Tech'': Pre-deployment Testing and Post-deployment Surveillance.}
Frontier models routinely undergo red-teaming and safety testing prior to public release; this typically covers threats such as biological, chemical, nuclear and cyber misuse, as well as qualitative checks around misalignment and deception. Psychological safety should be added to this list as a first-class release criterion: models should be benchmarked for sycophancy, excessive anthropomorphisation, and delusion reinforcement prior to release, with results published in model cards so that claims of safety are auditable rather than taken on trust.

Because pre-deployment testing cannot anticipate all real-world harm, post-release surveillance is equally necessary. Frontier technology companies should collaborate cross-functionally with clinicians and researchers to characterise the nature and trajectory of delusional and harmful conversations; on this basis it may become possible to detect high-risk conversational trajectories and steer users away from them. Candidate technical interventions span a spectrum of sophistication, from system prompt adjustments, through inference-time guardrails and classifiers and limits on session length and continuity, to deeper alterations to pre-training and preference-based fine-tuning. Each involves trade-offs between safety, utility, and user privacy that warrant empirical study rather than assumption. Developers should additionally provide accessible mechanisms for users and families to report harmful model behaviour, feeding the same surveillance loop. Provenance and accountability mechanisms, such as cryptographically signed outputs or secure conversation logging, could in principle support post-incident review of a model's role in a harmful act, analogous to a flight recorder in aviation. We note, however, that text watermarking remains technically fragile (it can be degraded by paraphrasing), and that any such logging must be balanced against significant privacy and confidentiality concerns, particularly given the sensitive nature of mental health-related conversations.

\textbf{Researchers and Academic Institutions: From Anecdote to Evidence.}
The research agenda spans both sides of the human--machine dyad. On the clinical side, priorities include establishing case registries and longitudinal cohorts to move the evidence base beyond anecdote, deriving a consensus case definition, and identifying predisposing and protective factors. On the technical side, priorities include developing systems to detect sycophantic and delusion-reinforcing model behaviour, refining benchmarks that measure the propensity of models to co-construct delusional content, and evaluating pre- and post-training mitigation methods. Crucially, these strands should not run in parallel but in dialogue: clinical phenotypes should inform benchmark design, and benchmark findings should generate hypotheses testable in clinical populations.

\textbf{Regulatory Bodies and Governmental Institutions: Closing the Loop.}
At a societal level, governments should consider carefully the effects of LLM-associated psychological destabilisation and influence. The wider spectrum of effect, from milder shifts in psychological tendencies and biases through to overt psychiatric disease, can not only harm population mental health but also drive division and polarisation, much like the documented effects of social media \cite{Kubin2021}. Regulators are the actor capable of making the surveillance loop mandatory rather than voluntary. Existing post-market surveillance infrastructure, such as the MHRA Yellow Card scheme, could be adapted to capture AI-related mental health harms, creating a formal reporting channel equivalent to that for adverse drug reactions \cite{MHRA2025YellowCard}. Reporting guidelines such as CONSORT-AI \cite{Liu2020CONSORTAI} could be expanded to explicitly cover sycophancy and psychological safety outcomes in trials of AI-based interventions. The December 2025 NAAG letter demonstrates that enforcement bodies are already prepared to treat sycophantic and delusional outputs as a consumer protection issue \cite{NAAG2025Letter}; formal regulatory frameworks should follow, specifying disclosure requirements, third-party audit rights, and incident notification duties.

In conclusion, whether or not AI-associated psychosis ultimately earns recognition as a distinct diagnostic entity, the harms it describes are real enough to warrant coordinated action now, and the nosological debate must not become a reason for delay. The phenomenon sits at the intersection of clinical medicine and machine learning, and so does its solution: clinicians who ask the right questions, researchers who convert cases into evidence, developers who treat psychological safety as a release criterion, and regulators who make surveillance and accountability mandatory. LLMs are the latest in a long line of dual-use technologies posing both benefit and harm to society; what is different this time is that the harm operates through conversation, with unique mechanisms of sycophancy and anthropomorphism. It is essential that all stakeholders are informed and take an active part in shaping the future of this technology.

\newpage

\bibliographystyle{unsrt}

\bibliography{references}

@article{AuYeung2025PsychogenicMachine,
  author        = {Au Yeung, J. and Dalmasso, J. and Foschini, L. and Dobson, R. J. and Kraljevic, Z.},
  title         = {The Psychogenic Machine: Simulating {AI} Psychosis, Delusion Reinforcement and Harm Enablement in Large Language Models},
  journal       = {arXiv preprint arXiv:2509.10970},
  year          = {2025},
  url           = {https://arxiv.org/abs/2509.10970},
  note = {Available from: \url{https://arxiv.org/abs/2509.10970}}
}

@article{Pierre2025YoureNotCrazy,
  author  = {Pierre, Joseph M. and Gaeta, Ben and Raghavan, Govind and Sarma, Karthik V.},
  title   = {{``You're Not Crazy'': A Case of New-onset AI-associated Psychosis}},
  journal = {Innovations in Clinical Neuroscience},
  year    = {2025},
  volume  = {22},
  number  = {10--12},
  pages   = {11--13}
}

@article{Ostergaard2023ChatbotsGenerateDelusions,
  author       = {{\O}stergaard, S{\o}ren D.},
  title        = {Will Generative Artificial Intelligence Chatbots Generate Delusions in Individuals Prone to Psychosis?},
  journal      = {Schizophrenia Bulletin},
  year         = {2023},
  volume       = {49},
  number       = {6},
  pages        = {1418--1419},
  doi          = {10.1093/schbul/sbad128}
}

@misc{Morrin2025DialsOfBelief,
  author       = {Morrin, Hamilton and Deeley, Quinton and Pollak, Thomas},
  title        = {Playing with the dials of belief: how controllable {AI} behaviours could modulate human belief and cognition across scales},
  howpublished = {PsyArXiv},
  year         = {2025},
  url          = {https://osf.io/preprints/psyarxiv/7qcv8_v1/},
  urldate      = {2025-12-30},
  doi          = {10.31234/osf.io/7qcv8_v1},
  note = {Available from: \url{https://osf.io/preprints/psyarxiv/7qcv8_v1/} [Accessed: 2025-12-30]}
}

@misc{Morrin2025DelusionsByDesign,
  author = {Morrin, Hamilton and Nicholls, Luke and Levin, Michael and Yiend, Jenny and Iyengar, Umesh and DelGuidice, Francesco and others},
  title = {Delusions by design? How everyday {AIs} might be fuelling psychosis (and what can be done about it)},
  howpublished = {OSF Preprints},
  year = {2025},
  month = {September},
  url = {https://osf.io/cmy7n_v5},
  urldate = {2025-09-25},
  doi = {10.31219/osf.io/cmy7n_v5},
  note = {Available from: \url{https://osf.io/cmy7n_v5} [Accessed: 2025-09-25]}
}

@article{Hudon2025JMIR,
  author = {Hudon, Alexandre and Stip, Emmanuel},
  title = {Delusional Experiences Emerging From {AI} Chatbot Interactions or ``{AI} Psychosis''},
  journal = {JMIR Mental Health},
  year = {2025},
  volume = {12},
  number = {1},
  pages = {e85799},
  month = {December},
  doi = {10.2196/85799}
}

@article{Kendell2003Validity,
  author = {Kendell, R. and Jablensky, A.},
  title = {Distinguishing between the validity and utility of psychiatric diagnoses},
  journal = {American Journal of Psychiatry},
  year = {2003},
  volume = {160},
  number = {1},
  pages = {4--12},
  doi = {10.1176/appi.ajp.160.1.4}
}

@article{Jablensky2016Validity,
  author = {Jablensky, A.},
  title = {Psychiatric classifications: validity and utility},
  journal = {World Psychiatry},
  year = {2016},
  volume = {15},
  number = {1},
  pages = {26--31},
  doi = {10.1002/wps.20284}
}

@article{sharma2023sycophancy,
  author        = {Sharma, Mrinank and Tong, Meg and Korbak, Tomasz and Duvenaud, David and Askell, Amanda and Bowman, Samuel R. and Cheng, Newton and Durmus, Esin and Hatfield-Dodds, Zac and Johnston, Scott R. and Kravec, Shauna and Maxwell, Timothy and McCandlish, Sam and Ndousse, Kamal and Rausch, Oliver and Schiefer, Nicholas and Yan, Da and Zhang, Miranda and Perez, Ethan},
  title         = {Towards Understanding Sycophancy in Language Models},
  journal       = {arXiv preprint arXiv:2310.13548},
  year          = {2023},
  eprint        = {2310.13548},
  archivePrefix = {arXiv},
  primaryClass  = {cs.CL},
  url           = {https://arxiv.org/abs/2310.13548},
  note = {Available from: \url{https://arxiv.org/abs/2310.13548}}
}

@misc{openai2025sensitive,
  author       = {{OpenAI}},
  title        = {Strengthening {ChatGPT}'s responses in sensitive conversations},
  year         = {2025},
  month        = oct,
  url          = {https://openai.com/index/strengthening-chatgpt-responses-in-sensitive-conversations/},
  urldate      = {2026-01-16},
  organization = {OpenAI},
  note = {Available from: \url{https://openai.com/index/strengthening-chatgpt-responses-in-sensitive-conversations/} [Accessed: 2026-01-16]}
}

@article{fanous2025syceval,
  title         = {SycEval: Evaluating {LLM} Sycophancy},
  journal       = {arXiv preprint arXiv:2502.08177},
  author        = {Aaron Fanous and Jacob Goldberg and Ank A. Agarwal and Joanna Lin and Anson Zhou and Roxana Daneshjou and Sanmi Koyejo},
  year          = {2025},
  eprint        = {2502.08177},
  archivePrefix = {arXiv},
  primaryClass  = {cs.AI},
  url           = {https://arxiv.org/abs/2502.08177},
  note = {Available from: \url{https://arxiv.org/abs/2502.08177}}
}

@misc{hong2025measuringsycophancylanguagemodels,
      title={Measuring Sycophancy of Language Models in Multi-turn Dialogues},
      author={Jiseung Hong and Grace Byun and Seungone Kim and Kai Shu and Jinho D. Choi},
      year={2025},
      eprint={2505.23840},
      archivePrefix={arXiv},
      primaryClass={cs.CL},
      url={https://arxiv.org/abs/2505.23840},
  note = {Available from: \url{https://arxiv.org/abs/2505.23840}}
}

@misc{yuan2025echobenchbenchmarkingsycophancymedical,
      title={EchoBench: Benchmarking Sycophancy in Medical Large Vision-Language Models},
      author={Botai Yuan and Yutian Zhou and Yingjie Wang and Fushuo Huo and Yongcheng Jing and Li Shen and Ying Wei and Zhiqi Shen and Ziwei Liu and Tianwei Zhang and Jie Yang and Dacheng Tao},
      year={2025},
      eprint={2509.20146},
      archivePrefix={arXiv},
      primaryClass={cs.CV},
      url={https://arxiv.org/abs/2509.20146},
  note = {Available from: \url{https://arxiv.org/abs/2509.20146}}
}

@article{Kubin2021,
    author = {Kubin, Emily and von Sikorski, Christian},
    title = {The Role of (Social) Media in Political Polarization: A Systematic Review},
    journal = {Annals of the International Communication Association},
    volume = {45},
    number = {3},
    pages = {188-206},
    year = {2021},
    month = {09},
    issn = {2380-8985},
    doi = {10.1080/23808985.2021.1976070},
    url = {https://doi.org/10.1080/23808985.2021.1976070}
}

@article{Enders2023SocialMedia,
  title   = {The Relationship Between Social Media Use and Beliefs in Conspiracy Theories and Misinformation},
  author  = {Enders, Adam M. and Uscinski, Joseph E. and Seelig, Michelle I. and Klofstad, Casey A. and Wuchty, Stefan and Funchion, John R. and Murthi, Manohar N. and Premaratne, Kamal and Stoler, Justin},
  journal = {Political Behavior},
  year    = {2023},
  volume  = {45},
  number  = {2},
  pages   = {781--804},
  doi     = {10.1007/s11109-021-09734-6},
  publisher = {Springer}
}

@article{auyeungchatbots2023,
  author  = {Au Yeung, Joshua and Kraljevic, Zeljko and Luintel, Akish and Balston, Alfred and Idowu, Esther and Dobson, Richard J. and Teo, James T.},
  title   = {{AI} chatbots not yet ready for clinical use},
  journal = {Frontiers in Digital Health},
  volume  = {5},
  year    = {2023},
  doi     = {10.3389/fdgth.2023.1161098}
}

@misc{OpenAIKenyaWorkers2023,
  author    = {Perrigo, Billy},
  title     = {{OpenAI} used {K}enyan workers on less than \$2 per hour to make {ChatGPT} less toxic},
  publisher = {TIME},
  year      = {2023},
  month     = jan,
  url       = {https://time.com/6247678/openai-chatgpt-kenya-workers/},
  urldate   = {2026-02-26},
  note = {Available from: \url{https://time.com/6247678/openai-chatgpt-kenya-workers/} [Accessed: 2026-02-26]}
}

@misc{schimmelpfennig2026humanlikeaidesignincreases,
title={Humanlike AI Design Increases Anthropomorphism but Yields Divergent Outcomes on Engagement and Trust Globally},
author={Robin Schimmelpfennig and Mark Díaz and Vinodkumar Prabhakaran and Aida Davani},
year={2026},
eprint={2512.17898},
archivePrefix={arXiv},
primaryClass={cs.AI},
url={https://arxiv.org/abs/2512.17898},
  note = {Available from: \url{https://arxiv.org/abs/2512.17898}}
}

@article{BrandtzaegAIfriend2022,
author = {Brandtzaeg, Petter Bae and Skjuve, Marita and Følstad, Asbjørn},
title = {My AI Friend: How Users of a Social Chatbot Understand Their Human–AI Friendship},
journal = {Human Communication Research},
volume = {48},
number = {3},
pages = {404-429},
year = {2022},
month = {04},
issn = {1468-2958},
doi = {10.1093/hcr/hqac008},
url = {https://doi.org/10.1093/hcr/hqac008}
}

@misc{cohn2024believinganthropomorphismexaminingrole,
title={Believing Anthropomorphism: Examining the Role of Anthropomorphic Cues on Trust in Large Language Models},
author={Michelle Cohn and Mahima Pushkarna and Gbolahan O. Olanubi and Joseph M. Moran and Daniel Padgett and Zion Mengesha and Courtney Heldreth},
year={2024},
eprint={2405.06079},
archivePrefix={arXiv},
primaryClass={cs.HC},
url={https://arxiv.org/abs/2405.06079},
  note = {Available from: \url{https://arxiv.org/abs/2405.06079}}
}

@article{Liu2020CONSORTAI,
author = {Liu, Xiaoxuan and Cruz Rivera, S. and Moher, D. and Calvert, M. J. and Denniston, A. K. and {SPIRIT-AI and CONSORT-AI Working Group}},
title = {Reporting guidelines for clinical trial reports for interventions involving artificial intelligence: the {CONSORT-AI} extension},
journal = {The Lancet Digital Health},
year = {2020},
volume = {2},
number = {10},
pages = {e537--e548},
month = {October},
doi = {10.1016/S2589-7500(20)30218-1},
pmid = {33328048}
}

@article{Hill2025NYT,
  author = {Hill, Kashmir},
  title = {They Asked an {A.I.} Chatbot Questions. {T}he Answers Sent Them Spiraling},
  journal = {The New York Times},
  year = {2025},
  month = {June},
  url = {https://www.nytimes.com/2025/06/13/technology/chatgpt-ai-chatbots-conspiracies.html},
  note = {Available from: \url{https://www.nytimes.com/2025/06/13/technology/chatgpt-ai-chatbots-conspiracies.html}}
}

@article{Hill2025NYTSpiral,
  author = {Hill, Kashmir and Freedman, D.},
  title = {Chatbots Can Go Into a Delusional Spiral. {H}ere's How It Happens},
  journal = {The New York Times},
  year = {2025},
  month = {August},
  day = {8},
  url = {https://www.nytimes.com/2025/08/08/technology/ai-chatbots-delusions-chatgpt.html},
  urldate = {2025-08-28},
  note = {Available from: \url{https://www.nytimes.com/2025/08/08/technology/ai-chatbots-delusions-chatgpt.html} [Accessed: 2025-08-28]}
}

@article{Hill2025NYTReality,
  author = {Hill, Kashmir and Valentino-DeVries, J.},
  title = {What {OpenAI} Did When {ChatGPT} Users Lost Touch With Reality},
  journal = {The New York Times},
  year = {2025},
  month = {November},
  day = {23},
  url = {https://www.nytimes.com/2025/11/23/technology/openai-chatgpt-users-risks.html},
  urldate = {2025-11-23},
  note = {Available from: \url{https://www.nytimes.com/2025/11/23/technology/openai-chatgpt-users-risks.html} [Accessed: 2025-11-23]}
}

@article{Klee2025RollingStone,
  author = {Klee, Miles},
  title = {People Are Losing Loved Ones to {AI}-Fueled Spiritual Fantasies},
  journal = {Rolling Stone},
  year = {2025},
  month = {May},
  day = {4},
  url = {https://www.rollingstone.com/culture/culture-features/ai-spiritual-delusions-destroying-human-relationships-1235330175/},
  note = {Available from: \url{https://www.rollingstone.com/culture/culture-features/ai-spiritual-delusions-destroying-human-relationships-1235330175/}}
}

@misc{Chatterjee2025NPR,
  author = {Chatterjee, Rhitu},
  title = {Their teenage sons died by suicide. Now, they are sounding an alarm about {AI} chatbots},
  organization = {NPR},
  year = {2025},
  month = {September},
  url = {https://www.npr.org/sections/shots-health-news/2025/09/19/nx-s1-5545749/ai-chatbots-safety-openai-meta-characterai-teens-suicide},
  urldate = {2025-12-30},
  note = {Available from: \url{https://www.npr.org/sections/shots-health-news/2025/09/19/nx-s1-5545749/ai-chatbots-safety-openai-meta-characterai-teens-suicide} [Accessed: 2025-12-30]}
}

@article{Ostergaard2025Mania,
  author = {{\O}stergaard, S. D.},
  title = {Emotion contagion through interaction with generative artificial intelligence chatbots may contribute to development and maintenance of mania},
  journal = {Acta Neuropsychiatrica},
  year = {2025},
  volume = {37},
  pages = {e79}
}

@article{Olsen2025medRxiv,
  author = {Olsen, S. G. and Reinecke-Tellefsen, C. J. and {\O}stergaard, S. D.},
  title = {Potentially harmful consequences of artificial intelligence ({AI}) chatbot use among patients with mental illness: Early data from a large psychiatric service system},
  journal = {medRxiv},
  year = {2025},
  pages = {2025.11.19.25340580},
  doi = {10.1101/2025.11.19.25340580},
  url = {https://www.medrxiv.org/content/10.1101/2025.11.19.25340580v1},
  note = {Preprint cited 2025 Nov 29. Available from: \url{https://www.medrxiv.org/content/10.1101/2025.11.19.25340580v1}}
}

@article{Darvesh2020GamingDisorder,
  author = {Darvesh, N. and Radhakrishnan, A. and Lachance, C. C. and Nincic, V. and Sharpe, J. P. and Ghassemi, M. and others},
  title = {Exploring the prevalence of gaming disorder and Internet gaming disorder: a rapid scoping review},
  journal = {Systematic Reviews},
  year = {2020},
  volume = {9},
  number = {1},
  pages = {68},
  month = {April},
  doi = {10.1186/s13643-020-01329-2}
}

@incollection{Roozenbeek2024Inoculation,
  author = {Roozenbeek, Jon and van der Linden, Sander},
  title = {Countering misinformation through psychological inoculation},
  booktitle = {Advances in Experimental Social Psychology},
  publisher = {Academic Press},
  year = {2024},
  volume = {69},
  pages = {1--58},
  doi = {10.1016/bs.aesp.2023.11.001},
  url = {https://www.sciencedirect.com/science/article/pii/S0065260123000266},
  note = {Accessed 2025-12-31. Available from: \url{https://www.sciencedirect.com/science/article/pii/S0065260123000266}}
}

@misc{MHRA2025YellowCard,
  author = {{Medicines and Healthcare products Regulatory Agency (MHRA)}},
  title = {Software, apps and Artificial Intelligence ({AI}) reporting | Making medicines and medical devices safer},
  year = {2025},
  url = {https://yellowcard.mhra.gov.uk/software-apps-and-artificial-intelligence},
  urldate = {2025-09-25},
  organization = {Yellow Card Scheme},
  note = {Available from: \url{https://yellowcard.mhra.gov.uk/software-apps-and-artificial-intelligence} [Accessed: 2025-09-25]}
}

@misc{NAAG2025Letter,
  author = {{National Association of Attorneys General}},
  title = {Letter to the legal representatives of {Anthropic}, {Apple}, {Chai AI}, {Character Technologies}, {Google}, {Luka}, {Meta}, {Microsoft}, {Nomi AI}, {OpenAI}, {Perplexity AI}, {Replika}, and {xAI}},
  year = {2025},
  url = {https://www.iowaattorneygeneral.gov/media/cms/12_68B5C629180F6.pdf},
  urldate = {2025-12-30},
  organization = {Iowa Attorney General's Office},
  note = {Available from: \url{https://www.iowaattorneygeneral.gov/media/cms/12_68B5C629180F6.pdf} [Accessed: 2025-12-30]}
}

@misc{OpenAI2025Grants,
  author = {{OpenAI}},
  title = {Funding grants for new research into {AI} and mental health},
  year = {2025},
  url = {https://openai.com/index/ai-mental-health-research-grants/},
  urldate = {2025-12-31},
  note = {Available from: \url{https://openai.com/index/ai-mental-health-research-grants/} [Accessed: 2025-12-31]}
}

@article{Dohnany2025FolieADeux,
  author = {Dohn{\'{a}}ny, S. and Kurth-Nelson, Z. and Spens, E. and Luettgau, L. and Reid, A. and Gabriel, I. and others},
  title = {Technological folie \`{a} deux: Feedback Loops Between {AI} Chatbots and Mental Illness},
  journal = {arXiv preprint arXiv:2507.19218},
  year = {2025},
  url = {http://arxiv.org/abs/2507.19218},
  note = {Accessed 2025-12-31. Available from: \url{http://arxiv.org/abs/2507.19218}}
}

@article{Vesterinen2020LoopingEffect,
  author = {Vesterinen, T.},
  title = {Identifying the Explanatory Domain of the Looping Effect: Congruent and Incongruent Feedback Mechanisms of Interactive Kinds},
  journal = {Journal of Social Ontology},
  year = {2020},
  volume = {6},
  number = {2},
  pages = {159--185},
  doi = {10.1515/jso-2020-0016}
}

@article{Hacking2006MakingUpPeople,
  author = {Hacking, Ian},
  title = {Making Up People},
  journal = {London Review of Books},
  year = {2006},
  volume = {28},
  number = {16},
  month = {August},
  url = {https://www.lrb.co.uk/the-paper/v28/n16/ian-hacking/making-up-people},
  urldate = {2025-12-31},
  note = {Available from: \url{https://www.lrb.co.uk/the-paper/v28/n16/ian-hacking/making-up-people} [Accessed: 2025-12-31]}
}

@article{Tsou2007Looping,
  author = {Tsou, J. Y.},
  title = {Hacking on the Looping Effects of Psychiatric Classifications: What Is an Interactive and Indifferent Kind?},
  journal = {International Studies in the Philosophy of Science},
  year = {2007},
  volume = {21},
  number = {3},
  pages = {329--344},
  doi = {10.1080/02698590701755104}
}

@book{apa2022dsm5tr,
  author    = {{American Psychiatric Association}},
  title     = {Diagnostic and Statistical Manual of Mental Disorders},
  edition   = {5th ed., text rev.},
  publisher = {American Psychiatric Association Publishing},
  address   = {Washington, DC},
  year      = {2022},
  doi       = {10.1176/appi.books.9780890425787}
}

@misc{who2019icd11,
  author    = {{World Health Organization}},
  title     = {International Statistical Classification of Diseases and Related Health Problems},
  edition   = {11th},
  year      = {2019},
  url       = {https://icd.who.int/},
  note      = {Available from: \url{https://icd.who.int/}}
}

@report{yougov2026aithinking,
  author       = {{YouGov}},
  title        = {AI Thinking -- YouGov Survey},
  year         = {2026},
  month        = {July},
  note         = {Fieldwork: July 16--20, 2026; N = 1,110 U.S. adults},
  url          = {https://ygo-assets-websites-editorial-emea.yougov.net/documents/AI_Thinking_poll_results.pdf},
  urldate      = {2026-08-18},
  type         = {Survey report},
  institution  = {YouGov}
}

@online{mind2026ai,
  author       = {{Mind}},
  title        = {Mind launches {AI} and {Mental Health} {Commission}},
  year         = {2026},
  month        = {feb},
  day          = {20},
  url          = {https://www.mind.org.uk/news-campaigns/news/mind-launches-ai-and-mental-health-commission/},
  urldate      = {2026-08-18},
  organization = {Mind},
  note         = {News announcement}
}

\end{document}